 \documentclass[final,5p,times,twocolumn,authoryear]{elsarticle}

\usepackage{amssymb}
\usepackage{amsmath}

\usepackage[colorlinks]{hyperref}
\hypersetup{linkcolor=red,citecolor=black,filecolor=cyan,urlcolor=blue}
\usepackage{bm}
\usepackage{xcolor}
\newcommand{\kms}{\mbox{\,km\,s$^{-1}$}\,}

 \usepackage{lineno}

\journal{ }

\begin{document}

\begin{frontmatter}



 \title{Reinterpretation of the Fermi acceleration of  cosmic rays  in terms of the ballistic surfing acceleration in supernova  shocks}


\author[label1,label2]{Krzysztof Stasiewicz} 
\affiliation[label1]{organization={Space Research Centre, Polish Academy of Sciences},
            addressline={Bartycka 18A}, 
            city={Warszawa},
            postcode={00-716}, 
            country={Poland}}
 \fntext[label2]{Corresponding author:  kstasiewicz@cbk.waw.pl}
\begin{abstract}
The applicability of first-order Fermi acceleration in explaining the cosmic ray spectrum has been reexamined using recent results on shock acceleration mechanisms from the Multiscale Magnetospheric mission in Earth's bow shock. It is demonstrated that the Fermi mechanism is a crude approximation of the ballistic surfing acceleration (BSA) mechanism. While both mechanisms yield similar expressions for the energy gain of a particle after encountering a shock once, leading to similar power-law distributions of the cosmic ray energy spectrum, the Fermi mechanism is found to be inconsistent with fundamental equations of electrodynamics.
  It is shown that the spectral index of cosmic rays is determined by the average magnetic field compression rather than the density compression, as in the Fermi model.   It is shown that the knee observed in the spectrum at an energy of $5\times10^{15}$\,eV could correspond to ions with a gyroradius comparable to the size of shocks in supernova remnants. The BSA mechanism can accurately reproduce the observed spectral index $s = -2.5$ below the knee energy, as well as a steeper spectrum, $s = -3$, above the knee. The acceleration time up to the knee, as implied by BSA, is on the order of 300 years.
 First-order Fermi acceleration does not represent a physically valid mechanism and should be replaced by ballistic surfing acceleration in  applications or models related to quasi-perpendicular shocks in space. It is noted that BSA, which operates outside of shocks, was previously misattributed to shock drift acceleration (SDA), which operates within shocks.

\end{abstract}


%


\end{frontmatter}




\section{Introduction}
Cosmic rays exhibit a power-law energy distribution with a spectral index of $s\approx -2.5$  below the knee at an energy of $K\approx 5{\times}10^{15}$\,eV,  and a steeper slope, $s\approx -3$,  above the knee and below the ankle at $5{\times}10^{18}$\,eV \citep{Hillas:1984,Helder:2012}. Understanding the spectrum and acceleration mechanisms operating at such high energies has been an important, yet  not fully explained problem in astrophysics.
  \citet{Fermi:1949} proposed that the cosmic ray spectrum could correspond to ions accelerated by bouncing off magnetic clouds in the interstellar medium. It is now believed that these clouds represent magnetic turbulence responsible for second-order Fermi acceleration. On the other hand, first-order Fermi acceleration is generally regarded as the major mechanism responsible for the formation of cosmic rays via the diffusive shock acceleration (DSA) process \citep{Bell:1978,Blandford:1987,Longair:2011}.  

The recent Magnetospheric Multiscale (MMS) mission \citep{Burch:2016}, consisting of four satellites flying through the Earth's bow shock, has provided the best measurements of collisionless shocks to date. The MMS satellites fly occasionally with separation distances of 20 km, which is smaller than the thermal proton gyroradius of 100 km. In this mission the electric, $\bm{E}$ and magnetic, $\bm{B}$ fields are sampled at the rate of 8192 s$^{-1}$, while particles distribution functions are measured with time  resolution of 30\,ms  for electrons and 150\,ms for ions. 

Based on MMS data, it has been established that thermalization, heating, and acceleration of ions and electrons in collisionless shocks are related to four plasma processes: SWE (stochastic wave energization), TTT (transit time thermalization), BSA (ballistic surfing acceleration), and QAH (quasi-adiabatic heating), explained further in the text (see also list of acronyms). Understanding processes in the bow shock can help understand more powerful astrophysical shocks, such as those created during supernova explosions, which are most likely involved in the acceleration of cosmic rays.

The processes SWE  and TTT  are stochastic in nature, related to deterministic chaos. They rely on strong gradients of the electric and magnetic fields that lead to randomization of particle orbits and efficient stochastic heating on the timescale of one gyroperiod \citep{Stasiewicz:2023MN2}. TTT thermalizes streaming ions on magnetic field gradients within a fraction of the gyroperiod and does not require any waves, instabilities, or anomalous collisions. SWE works on electric field gradients and can accelerate protons to a few hundred keV \citep{Stasiewicz:2021MN,Stasiewicz:2021JR,Stasiewicz:2022MN}. SWE of electrons is responsible both for heating and formation of flat-top electron distributions \citep{Stasiewicz:2023MN3}.
Quasi-adiabatic heating (QAH) operates on particles with a conserved first adiabatic invariant ($v_\perp^2/B=\mathrm{constant}$), which requires the gyroradius $r_c$ to be much smaller than the width of the shock, $D\equiv B|\bm{\nabla} B|^{-1}$.

Ballistic surfing acceleration (BSA) operates on particles with a gyroradius $r_c\gg D$, which applies to superthermal ions and mildly relativistic electrons \citep{Stasiewicz:2023MN2}.  It bears resemblance to the shock drift acceleration (SDA)  previously discussed by various authors \citep{Jokipii:1982, Jones:1991, Zank:1996}.  
Another similar process is shock surfing acceleration (SSA), where low energy particles drift along the shock front due to surface waves \citep{Shapiro:2001,Lever:2001,Hoshino:2002}. In all these processes, energization is due to particle motion along the convection electric field. However, the term SDA  implies  $\bm{\nabla} B$ drift acceleration, which occurs within the  ramp. In contrast, BSA occurs outside the ramp, where particles with large gyroradii engage in ballistic surfing and do not experience $\bm{\nabla} B$ or wave effects.

As is well known, plasma can also be  effectively heated by waves at resonance frequencies: cyclotron, lower hybrid, upper hybrid, and plasma frequencies. Resonant heating is commonly used in laboratory plasmas, but does not appear to be important in shocks.

   A question arises, what is the relation of the Fermi mechanism and the DSA process with the above-described processes identified at the bow shock. 
In the next section we shall study  heating  of energetic ions in a model shock,  searching for signatures of  Fermi acceleration. We find  that the first order Fermi acceleration represents  a crude  approximation of BSA. We show that the cosmic ray spectrum and the presence of the knee  can be explained exclusively  by BSA.

\section{Ballistic surfing acceleration}
To understand acceleration of high energy particles  we consider  motion of  test particles with  velocity $v$ much larger than the thermal velocity $v_{Ti}$ of ions that maintain the shock. Trajectories of particles with rest mass $m_0$ and charge $q$ are described by the momentum equation
\begin{equation}
 \frac{\mathrm{d}\bm{p}}{\mathrm{d}t}=q(\bm{E}+ \bm{v}\times \bm{B}),  \label{MM}
 \end{equation} 
 where $\bm{p}=\gamma m_0\bm{v},\; \gamma=(1-v^2/c^2)^{-1/2}$, and $c$ is the speed of light. Here, we utilize the shock reference frame  in a geometry where the $\hat{\bm{x}}$ axis is in the negative direction to the shock normal, $\hat{\bm{y}}$ is in the direction of the convection electric field, and the magnetic field is in the $[B_{x0},0,B_z(x)]$ plane.
 The  kinetic energy increase of a  particle implied by this equation is
\begin{equation}
\Delta K= q\int \bm{E}\cdot \bm{v}\, dt.  \label{DKE}
\end{equation}
There are three types of electric fields that determine particle dynamics and  energization processes in the shock frame: $\hat{\bm{y}}E_y$ -- the convection field, constant across one-dimensional structures, $\hat{\bm{x}}E_S(x)$ -- the cross-shock electric field, maintained by the electron pressure gradient, and $\tilde{\bm{E}}(\bm{r},t)$ -- the wave electric field.    In the reference frame moving with the convection velocity, $E_y$ vanishes, but  acceleration is facilitated by the inductive electric field, $\bm{\nabla} \times \bm{E}=-\partial \bm{B}/\partial t$ observed by a particle moving in a time varying magnetic field. SWE works best on thermal particles and is ineffective for high-speed particles, so we exclude electrostatic waves from  consideration in this paper.

 The model shock is described in \citet{Stasiewicz:2023MN2} and represents a magnetic ramp   with compression $c_B=B_d/B_u$ between downstream and upstream values. A normal component $B_{x0}$ implies the upstream field angle  $\cos\eta=B_{x0}/B_u$ from the normal direction.    Magnetic turbulence is not included in the shock model, but it can be easily implemented  similarly to electrostatic waves. It is expected to  lead to the isotropisation of particle distributions and could also scatter particles back to the shock.

Parameters that enter the model are as follows: the upstream sonic Mach number, {$M=V_{u}/v_{Ti}=8$}, the ratio of the thermal ion  gyroradius $r_{ci}$ to the width of the shock ramp, $r_{ci}/D=1$, and compression $c_B=4$. The transit time thermalization parameter  is  set to $\chi_B=8$ and the stochastic parameter for the cross-shock electric field to $\chi_S=1.5$;  see \citet{Stasiewicz:2023MN2} for definitions.

\begin{figure}[t]
\centering
\includegraphics[width=8.6cm]{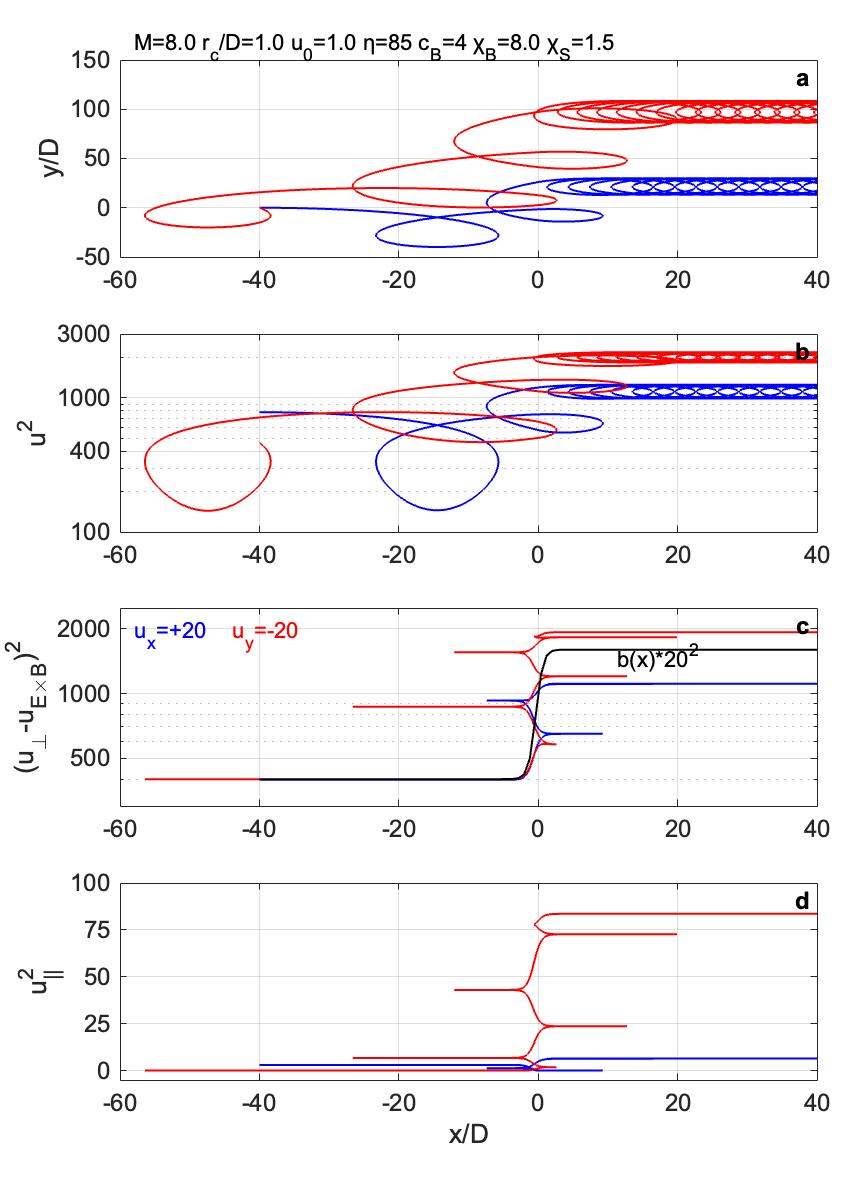}
\caption{Ion trajectories and heating by TTT and BSA in an oblique shock with a shock angle of  $\eta=85^\circ,\; \chi_{B}=8$, the cross-shock electric field $\chi_S=1.5$,  compression $c_B=4$, and thickness ratio $r_{ci}/D=1$, without waves. (a): Trajectories of two ions injected at $x=-40$ with a sonic Mach number $M=8$ and additional  velocities: $u_x=+20$ (blue) and $u_y=-20$ (red)  in units of the upstream ion thermal velocity $v_{Ti}$. (b):  Total kinetic energies of ions along the respective trajectories.  (c): Thermal (gyration) energies of ions along trajectories. The black curve shows theoretical adiabatic heating for the shock profile given by $b(x)\equiv B(x)/B_u$.  (d): Parallel energies of ions along trajectories. Typical values upstream of the bow shock are: $T_i\approx 20$\,eV, $v_{Ti}\approx 60$\kms, $r_{ci}\approx 100$\,km, $B_u=5$\,nT. \label{Ff1}}
\end{figure}

Two ions are injected into an oblique shock $\eta=85^\circ$ at $x=-40$  in units of the shock width  and followed with differential equations described in  \citet{Stasiewicz:2023MN2}. Fig.~\ref{Ff1}(a) shows  ion  trajectories in  plane $(x,y)$. The initial velocity consists of the  $E{\times}B$ drift normalized by the upstream thermal speed, $\bm{u}_{E{\times}B}=\bm{V}_{E{\times}B}/v_{Ti}$, with additional  velocities: $u_x=+20$ (blue), and  $u_y=-20$ (red). 
Panel (b) shows the total kinetic energies $u^2$ of ions along the respective trajectories,  panel (c)  shows gyration energies only $(\bm{u}_\perp-\bm{u}_{E{\times}B})^2$, and panel (d) shows $u_\parallel^2$.
Adiabatic projection of the initial perpendicular energy of the red ion is shown  as the black curve $b(x)u_{0\perp}^2$, where $b(x)=B(x)/B_u$.  Both ions behave non-adiabatically and do not follow the adiabatic projection. 

We observe that the blue ion makes three crossings of the shock at $x=0$, while the red ion makes seven crossings before being transmitted downstream. Each crossing is associated with an increase in gyration energy, as seen in panel (c). The kinetic energy in panel (b) correlates with particle movements in the $\pm y$ direction, along $E_y$, which has been defined as the BSA by \citet{Stasiewicz:2023MN2}.
This obvious acceleration process operating outside the shock was previously misattributed to SDA, which is related to $\bm{\nabla}B$ drift and operates within the shock. However, particles undergoing BSA in the gradient-free zone are unaware of the existence of the shock, and therefore cannot experience shock drift acceleration.  

Temperature variations in panel (c) and the transfer of energy into the parallel motion shown in panel (d) occur only during shock transitions at $x\approx 0$ and are caused by TTT, which is related to the inductive electric field or to the changing magnetic field direction in the shock frame. 

\begin{figure}[t]
\centering
\includegraphics[width=8.6cm]{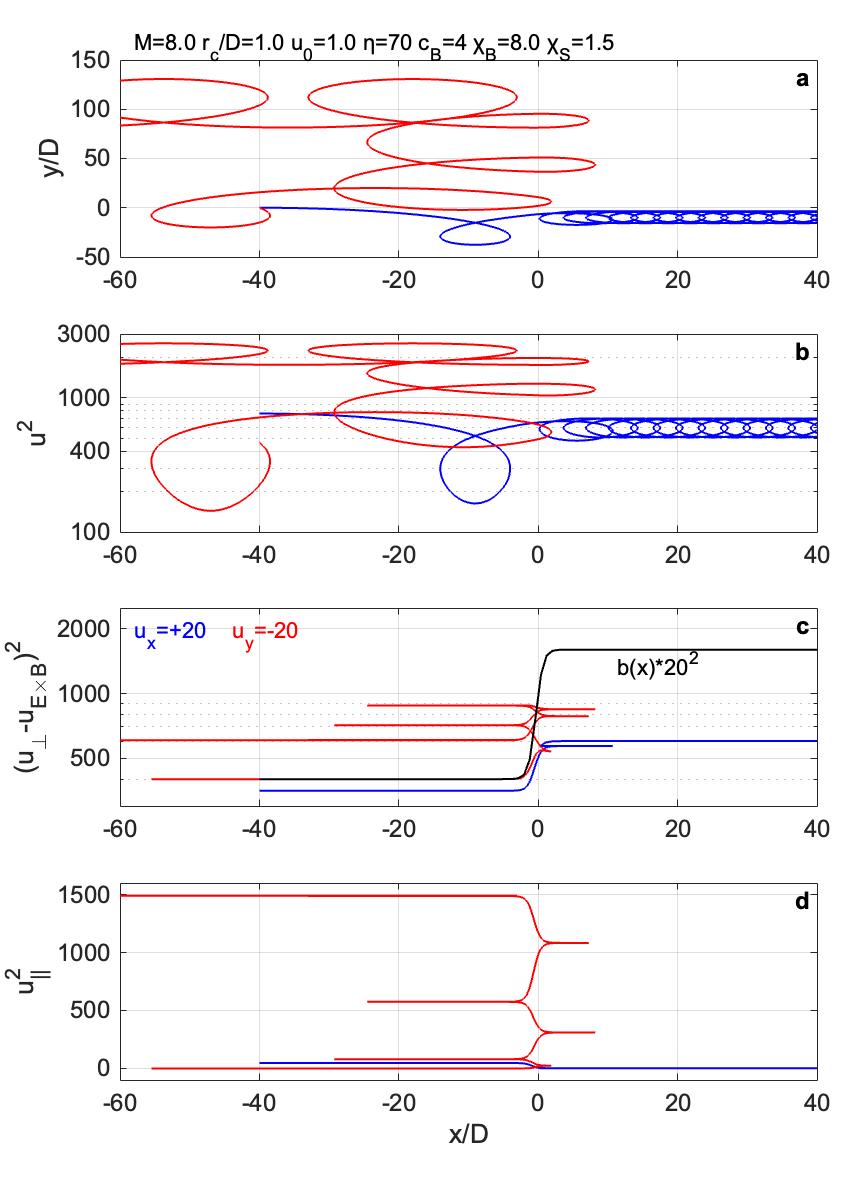}
\caption{BSA and TTT of ions in the same format  as in Fig.~\ref{Ff1} but for a shock angle $\eta=70^\circ$. The red ion is reflected upstream. Ion gyration energy is $u_\perp^2=400$ and the drift energy $M^2=64$ at the starting position $x=-40$. \label{Ff2}}
\end{figure}

By decreasing the shock angle to $\eta=70^\circ$ in Fig.~\ref{Ff2} we observe the reflection of the red ion back into the upstream region.
 Particle reflections are more common for smaller angles $\eta$, and are not related to the magnetic mirror force, which does not apply for large gyroradius particles, $r_c\gg D$.

For particles reflected upstream, the energy gain is directed into parallel motion rather than perpendicular heating. The transfer of energy into parallel motion, as shown in panels (d), is not related to the cross-shock electric field, which has a parallel component $E_\parallel=E_S(x)\cos\eta$, as the patterns remain the same when $E_S$ is set to zero.
The cross-shock potential is $e\Delta\Phi\approx 2(c_B-1)=6$ in normalized energy units of $T_{eu}\sim T_{iu}$, so it has negligible effect even for zero temperature ions with energy $u^2=M^2=64$; see \citet{Stasiewicz:2023MN2}, not to mention the high-energy ions, $u^2\sim 1000$ in the simulations here.  Furthermore, the cross-shock potential energy does not accumulate, but cancels out after each gyration around the shock.

One should not confuse cyclotron turning points with shock reflections. Cyclotron turning points recur periodically, once during every gyroperiod, while shock reflections are singular events that occur only once. 
Turning points can manifest anywhere, contingent upon the initial conditions and the gyroradius. Sensitivity to initial conditions is a hallmark of deterministic chaos. Figures \ref{Ff1}–\ref{Ff2}b illustrate that neither the process of reflection nor the transitions through the shock at $x\approx 0$ are associated with significant changes in particle energy.

\begin{figure}
\centering
\includegraphics[width=8.6cm]{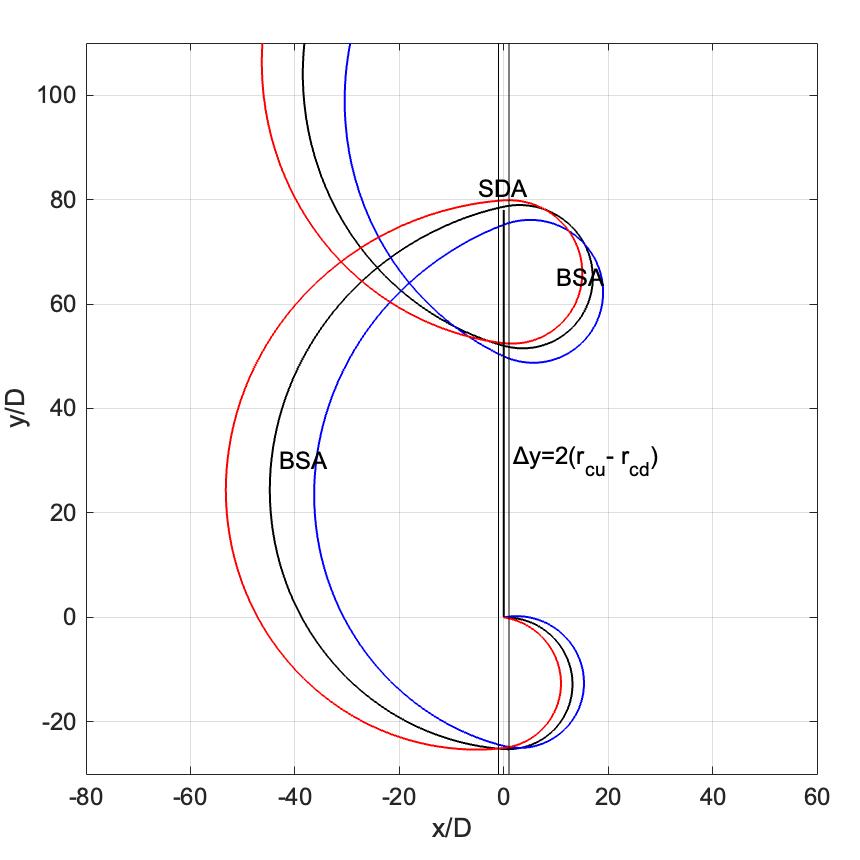}
\caption{The explanation of ballistic surfing acceleration: Three ions with a perpendicular velocity of $v_\perp=50v_{Ti}$ are injected at incident angles of $0^\circ$ (black), $+10^\circ$ (blue), and $-10^\circ$ (red) into the perpendicular shock located at position $x=0$. The parameters are set as follows: $c_B=4$, Mach number $M=3,\; r_{ci}/D=1$, and the cross-shock electric field $\chi_S=1.5$. The energy gain after one gyration across the shock is  $\Delta K\approx 2qE_y(r_{cu}-r_{cd})$, which leads directly to Eq.~(\ref{DK}).  SDA occurs within the ramp  denoted by vertical lines, where $|x|/D < 1$, while BSA operates in the gradient-free zone for $|x|/D \geq 1$. \label{Ff3}}
\end{figure}

\section{BSA  and formation of the cosmic ray spectrum}

As observed in the previous section, energetic particles are accelerated outside shocks through the ballistic surfing acceleration process by the convection electric field $E_y$, rather than through reflections as suggested by \citet{Fermi:1949}.  A necessary condition for BSA is that the particle's gyroradius is larger than the shock width.  The  BSA process is elucidated in Fig.~\ref{Ff3}, where it can be observed that ions moving in the $+y$ direction in the upstream part of the orbit increase kinetic energy ($qE_yv_y>0$), while those moving in the $-y$ direction in the downstream orbit decrease energy ($qE_yv_y<0$).  Because $B_d>B_u$, the downstream gyroradius $r_{cd}$ is smaller than the upstream one $r_{cu}$, and particles always gain energy. Within the shock ramp denoted by vertical lines, the particles experience gyrocenter drifts induced by $\bm{\nabla}B$  and are subject to the combined convection and cross-shock electric fields, $\hat{\bm{y}}E_y +\hat{\bm{x}}E_S(x)$.  The energy gain after a full rotation around the shock consists of three terms:
 \begin{equation}
 \Delta K=\int_{|x|\ge D}^{(\mathrm{BSA})} qE_y v_y dt + \int_{|x|<D}^{(\mathrm{SDA})} qE_y v_y dt  + \int_{|x|<D} qE_S  v_x dt \label{DBSA}
 \end{equation}
where the first term corresponds to ballistic surfing acceleration  outside the shock ramp, the second term to shock drift acceleration  within the  ramp, and the third term to acceleration by the cross-shock electric field. The first BSA term is the only significant one for high-energy particles because the integration domain is much larger than in the other two integrals ($2\pi r_c \gg 2D$).  It can be concluded that BSA is the primary mechanism for accelerating particles with large gyroradii ($r_c\gg D$), while the effects of SDA are negligible due to the very short interaction time compared to the gyroperiod. Conversely, for small gyroradii particles ($r_c\ll D$), SDA serves as the primary mechanism for the acceleration  and BSA does not apply. However, this process is equivalent to adiabatic heating, so SDA can be considered a subset of the more general QAH \citep{Stasiewicz:2023MN1}.

After one gyration across the shock the energy gain implied by Eq.~(\ref{DBSA}) is $\Delta K\approx 2qE_y(r_{cu}-r_{cd})$, as illustrated in Fig.~\ref{Ff3}. For  relativistic particles with  kinetic energy $K\approx pc$ and gyroradius $ r_c=p_\perp/qB$,  the energy gain is
\begin{equation}
\Delta K \approx 2qE_y\frac{1}{\pi}\int_0^\pi(r_{cu}-r_{cd})\sin\theta\,d\theta = K \frac{(c_B-1)}{c_B}\frac{V_u}{c}   \label{DK}
\end{equation}
where $V_u= E_y/B_u$ is the upstream convection velocity,  and the integral is the average over pitch angles, assuming isotropic distribution.  Averaging over incident angles of particles entering the shock would introduce a numerical factor $g\lesssim 1$ to this equation; see Fig.~\ref{Ff3}. Such a factor could be incorporated into a slightly modified value of $c_B$, which is a free parameter of the model.  The effective compression will be then $c_B^\prime=c_B/[c_B-g(c_B-1)]$.

 Equation (\ref{DK}) is analogous to the expression derived through a wholly distinct approach by \citet{Bell:1978}. This  formulation is commonly referred to as the first-order Fermi acceleration:
\begin{equation}
\Delta K_F = K \frac{(c_N - 1)}{3} \frac{V_u}{c}. \label{DKF}
\end{equation}
This equation was derived from the difference in particle energy between two inertial frames, characterized by the velocity ratio $V_u/V_d = n_d/n_u \equiv c_N$; see  \citet{Longair:2011} for derivation. Here, $c_N$ denotes the  density compression, typically set to the standard value  $c_N = 4$. It's important to note that this equation was mistakenly linked to energization. However, the energy difference between two inertial frames is simply a scalar value derived from the Lorentz transformation and doesn't relate to acceleration, which is defined by Eq.~(\ref{DKE}), and must be computed within the same reference frame. This yields the accurate expression (\ref{DK}), which depends on $c_B$ but remains independent of $c_N$.
Figs.~\ref{Ff1}-\ref{Ff2}b show that particles undergo energy gain over an extended period by ballistic surfing in the upstream region and lose energy in the downstream region, as described by $\mathrm{d}K/\mathrm{d}t=qE_yv_y$. Contrary to suppositions of the Fermi model, shock transitions, $|x|<D$, are not associated with significant energy changes, as can be seen in the aforementioned figures and in Eq.~(\ref{DBSA}).

 Cumulative energy increases  caused by multiple encounters of particles with shocks, combined with scattering by  turbulence outside the ramp represent diffusion in  velocity space,  described as DSA. 
  Because the BSA energy increase in Eq.~(\ref{DK}) is similar to the first-order Fermi acceleration (\ref{DKF}), we can follow  the standard approach to DSA, as described, for example, by \citet{Bell:1978} and  \citet{Longair:2011}.
Equation~(\ref{DK}) implies that the particle energy after one interaction  with a shock is
\begin{equation}
K_1=h K_{0};\quad   h=1+(1-1/c_B)V_u/c .  \label{KN}
\end{equation}
This acceleration is not related to the reflection or to the shock crossing process but  to the full gyration across the shock ramp with $c_B>1$.
 Let $P$ be the probability that  particles remain in the shock region after one interaction or gyration. Then, after $j$ interactions there are $N=N_0P^j$ particles with energies $K=K_0h^j$.
Eliminating $j$, one obtains
$N/N_0=(K/K_0)^{\ln P/\ln h}$, 
where $N$ is the number of particles that reached energy $K$ and can be accelerated further.  Using  \citet{Longair:2011} value for probability, $ P \approx 1-V_u/c$, we find the  spectrum of cosmic rays predicted by the BSA model 
\begin{equation}
\frac{\mathrm{d}N}{\mathrm{d}K}  \propto   K^{s}    \label{NK2}
\end{equation}
where
\begin{equation}
  s= \frac{\ln P}{\ln h}-1 = -\frac{2c_B-1}{c_B-1}  \label{SS}
  \end{equation}
which can be compared with the spectral index  predicted by the Fermi/DSA model \citep{Longair:2011} 
\begin{equation}
 s_F=\frac{\ln P}{\ln h_F}-1 = -\frac{c_N+2}{c_N-1} .    \label{SF}
\end{equation}
In deriving this equation,  $h_F= 1+\frac{1}{3}(c_N-1)V_u/c$ was used from incorrect Eq.~(\ref{DKF}). The Fermi/DSA  formulation neglects the impact of the electric field, which is crucial for particle energization. Moreover, it introduces an erroneous dependence on density compression, inconsistent with the fundamental equations (\ref{MM}) and (\ref{DKE}) which determine heating, but remain independent of plasma density.  Eq.~(\ref{DKF}) is clearly non-physical, because it diverges for large values of $c_N$, while the numerical factor in the correct Eq.~(\ref{DK}) approaches unity for $c_B\gg 1$.

While not being physically valid, Eq.~(\ref{SF}) provides one correct value $s_F=-2.5$ for $c_N=3$, coinciding with Eq.~(\ref{SS}) for $c_B=3$, and both agree with the measured  index $s$. This single, incidentally correct value of the Fermi/DSA model can possibly explain its success and popularity despite erroneous physical assumptions. The error has not been disclosed earlier, probably due to the observed relation $n\propto B$ in shocks, which yields $c_N\approx c_B$. The physical dependence of particle energy gain on $c_B$ has been confused with a circumstantial dependence on $c_N$.

The BSA  model would reproduce  the observed index $s\approx -2.5$  for the effective shock compression $c_B\approx 3$. 
 For the standard value  $c_B=4$, implied from Rankine-Hugoniot relations  \citep{Kennel:1988},  we obtain $s= -2.3$. Smaller compressions $c_B=2$ would lead to $s=-3$ observed above the knee, while magnetic walls, $c_B\gg 1$ would lead to the asymptotic  value $s=-2$.

Although BSA would function effectively at arbitrarily high energy, the acceleration in the upstream region will be counteracted by deceleration in the downstream region as the diameter of the orbit on the compressed side approaches the shock length $L$, see Fig.~\ref{Ff3}. The condition $2r_{cd} \sim L$ would result in a knee in the spectrum located at energy
\begin{equation}
K_{L} \sim \frac{qc}{2} \langle L B_d\rangle . \label{KL}
\end{equation}
Interestingly, this does not depend on shock velocity or particle mass. The observed distribution of supernova remnant sizes, $L_\mathrm{SNR}$, ranges from 1 pc ($= 3 \times 10^{16}$ m) to 200 pc \citep{Badenes:2010}. The observed knee energy, $K_{L} \approx 5 \times 10^{15}$ eV, is derived from Eq.~(\ref{KL}) for $\langle LB_d \rangle \sim 1$ nT pc, which could correspond to, for example, $B_d \approx 1$ nT on the inner (compressed) side of a spherically expanding supernova shock and $L \sim 1$ pc.
Shocks with length $L < 1$ pc inside supernova remnants with an effective compression $c_B = 3$ would lead to $s = -2.5$, which could explain the energy spectrum below the knee.
  
 However, even particles with a gyroradius much larger than the shock length can undergo further acceleration. This scenario may occur when downstream flows become stagnant at some distance from the shock, with $V_d \sim 0$, leading to the vanishing of the electric field in this region, $E_{yd} \sim 0$. Particles gyrating in the downstream region would not experience deceleration, as depicted in Figs.~\ref{Ff1}-\ref{Ff2}b, but they would still be subject to acceleration in the upstream region, $\mathrm{d}K/\mathrm{d}t=qE_yv_y>0$. This process could yield cosmic ray energies up to the ankle at $5 \times 10^{18}$ eV in shocks shorter than the gyroradius, $L < r_c$. The spectral index $s\approx -3$ in this energy range can be achieved with $c_B\approx 2$; however, other factors may also play a significant role in determining  the observed index.

\section{Discussion}
 BSA   requires  that the gyroradius of particles is larger than the shock width, $r_c/D>1$, which is fulfilled at the bow shock by protons with energy higher than 100\,eV and by electrons with energy higher than 180\,keV. 
Initial heating of cold  ions in shocks is accomplished by  TTT and SWE  mechanisms, which can be seen in Fig.~4 of \cite{Stasiewicz:2023MN2}, where  streaming  protons  with  temperature  $T_i\approx 20$\,eV,    are TTT thermalized and SWE energized  to 400\,$T_i=8$\,keV within a   gyroperiod.  In quasi-parallel shocks ions are accelerated by the SWE mechanism to $\sim$\,200\,keV which corresponds to the $E{\times}B$ velocity in the wave electric field, $V_{E{\times}B}=\tilde{E}_\perp/B$ with $\tilde{E}_\perp \sim 100$\,mV\,m$^{-1}$ being much larger than the convection field $E_y\sim 5$\,mV\,m$^{-1}$ \citep{Stasiewicz:2021MN,Stasiewicz:2021JR}. Further acceleration of these ions could continue by means of BSA  in subsequent shock encounters. 

 Let us assume that the injection energy for  BSA  is $K_0=10$\,keV, and the shock parameters are $c_B=4$, and  $V_u=10,000$\,\kms. Using the exact form of Eq.~(\ref{DK}) with the kinetic energy defined as $K\equiv (m_0^2c^4+p^2c^2)^{1/2}-m_0c^2$,  we find that protons reach the knee at $K{=5\times}10^{15}$\,eV after 657   BSA interactions without energy losses. In case of faster shock speeds, $V_u=20,000$\kms the final energy  would be reached after 334 BSA, while  $V_u=40,000$\kms requires only 172 BSA interactions. We have assumed here that transmitted particles could be scattered back to the shock by downstream turbulence, while particles reflected upstream can encounter another shock front, or can be turned back by upstream waves.

The physical picture of acceleration in the shock reference frame adopted in this paper is easier to understand than the traditional (Fermi) analysis of ion acceleration in the plasma reference frame, where the convection electric field is removed.  This approach ignores the fundamental laws of physics that particle trajectories  are determined by the Lorentz force equation (\ref{MM}), as illustrated in Figs.~\ref{Ff1}-\ref{Ff2}, and the energization is determined  by  Eq.~(\ref{DKE}), and not by the energy transformation between two inertial systems as in the Fermi/DSA model.    
Furthermore,  \citet{Fermi:1949} relies on the concept of  reflections by the magnetic mirror force, which is valid for particles with $r_c <D$, and  clearly not applicable for high-energy, large gyroradius particles. 

\section{Conclusions}
It is shown that ions and electrons with gyroradius $r_c\gg D$  are  accelerated  by the convection electric field $E_y$ in a process described as ballistic surfing acceleration. 
 BSA operates outside of shocks on particles  from approximately 100\,eV for protons up to very high energies observed in the cosmic ray spectrum.
BSA predicts  a knee in the spectrum when the gyroradius becomes comparable to the size of the shock, which determines the knee energy  given  by Eq.~(\ref{KL}).  The spectral index  in BSA is determined by the effective shock compression $c_B$, and can accurately reproduce  the observed index $s\approx-2.5$  below the knee energy as well as a steeper spectrum, $s\approx -3$, above the knee.  It is demonstrated that the Fermi/DSA model, which yields the spectral index (\ref{SF}), is inconsistent with the fundamental equations of electrodynamics (\ref{MM}) and (\ref{DKE}), and is therefore not valid. The popularity of the Fermi model  was unjustified, primarily due to the coincidental agreement of one spectral index value with the correct BSA model and with observations.

  It is found that to reach the knee energy  of 5$\times$10$^{15}$\,eV, a proton starting from 10~keV in a collisionless environment needs only 334 BSA interactions  in shocks moving with velocity $V_u=20,000$\kms.  
For  protons moving in the average magnetic field of $\langle B\rangle=$\,1\,nT, the net acceleration time of 334 gyroperiods would correspond   to  227 years  of sidereal time,  accounting for time dilation during each  gyroperiod.\\



\noindent {\bf List of Acronyms}\\
\noindent BSA -- ballistic surfing acceleration\\
\noindent DSA -- diffusive shock acceleration\\
\noindent QAH -- quasi adiabatic heating\\
\noindent SDA -- shock drift acceleration (a subset of QAH)\\
\noindent SSA --  shock surfing acceleration\\
\noindent SWE --  stochastic wave energization\\
\noindent TTT -- transit time thermalization\\


\section*{Data/software availability}
The mathematical shock model used to make Figs.~\ref{Ff1}-\ref{Ff3} was published in:  https://doi.org/10.1093/mnrasl/slad071.

\section*{Acknowledgements}
 This work has been supported by  Narodowe Centrum Nauki (NCN), Poland, through grant No. 2021/41/B/ST10/00823.

  \bibliographystyle{elsarticle-harv} 



\end{document}